\input amstex
\documentstyle{amsppt}

\def\doublespace{\advance\baselineskip by 12pt}

\pagewidth{35pc}
\pageheight{8.4in}
\hcorrection{1.5pc}
\loadbold

\hyphenation{amsppt}

\define\AMS{American Mathematical Society}
\hfuzz1pc 
\overfullrule=0pt

\define\thismonth{\ifcase\month 
  \or January\or February\or March\or April\or May\or June%
  \or July\or August\or September\or October\or November%
  \or December\fi}


\font\fourtn=cmr10 scaled \magstep2
\font\fourtnsy=cmsy10 scaled \magstep3
\font\fourtnbf=cmbx10 scaled \magstep3
\textfont2=\fourtnsy

\shipout\vbox to\vsize{%
\parindent=0pt 
\vskip5pc
\centerline{\fourtnbf Stochastic Simulation of The}
\medskip
\centerline{\fourtnbf Three Dimensional Quantum Vacuum}
\bigskip

\medskip
\rightline{\fourtn  February 1995}
\vfill
\tenpoint

This publication was typeset using \AmSTeX{}, the American
Mathematical\newline \quad Society's \TeX{} macro system.

\medskip
\begingroup\obeylines
\TeX{} is a trademark of the \AMS{}.
\endgroup
}


\topmatter
\title\nofrills Stochastic Simulation of The Three Dimensional Quantum Vacuum
\endtitle
\author { James G. Gilson} \endauthor
\affil { School of Mathematical Sciences,
             Queen Mary and Westfield College,
             Mile End Road,
             London E1 4NS,
             United Kingdom.}\endaffil
\address { School of Mathematical Sciences,
             Queen Mary and Westfield College,
             Mile End Road,
             London E1 4NS,
             United Kingdom.}\endaddress
\email {j.g.gilson\@uk.ac.qmw.maths} \endemail

\thanks PACS index categories: 03.65.Bz 12.90+b 02.30.Dk 02.50.Ey\endthanks

\abstract  A complete solution to the long standing problem of basing Schr\"odinger
quantum theory on standard stochastic theory is given. The solution covers all
{\it single\/} particle three-dimensional Schr\"odinger theory linear or nonlinear
and with any external potential present. The system is classical, set in a six-dimensional
space and involves vacuum polarization as the background process. Basic vacuum polarization
energy characterised oscillators are identified and then in assemblies are
analysed in terms of energy occurrence frequencies. The orbits of polarization
monopoles are given and shown to be elliptical on subspaces surfaces.
The basic process takes place at the speed of light and is of a statistical
{\it zitterbewegung\/} character. The orthodox quantum probability density bilinear
quadratic form is derived from angular momentum consideration within the system
which is shown to be a generalisation of the usual quantum structure. Two statistical
assemblies are identified, a linear one associated with superposition of eigenfunctions
and a quadratic one associated with interactions between eigenstates. It is suggested
that this two-tier probabilistic system will remove some possible paradoxes that
plague the orthodox thoory. The relation of vacuum polarization in this work with its
occurrence in other physical contexts and a connection with {\it spin\/} is discussed.
 \endabstract
\endtopmatter
\document

\head 1. Introduction \endhead

In a series of earlier papers [12,13,14,15], the author has 
demonstrated that Schr\"odinger quantum theory can be alternatively 
expressed and constructed from classical physical concepts provided 
that the {\it dimensionality\/} of the background configuration space 
involved is increased and use is made of the idea of {\it vacuum\/} 
polarisation to explain and build the structure of the theory. These 
ideas were then taken further in reference [16], where it was shown 
that such an alternative theory based on classical {\it physical\/}
concepts can be reformulated from a more fundamental level to 
solve the long standing problem of founding Schr\"odinger theory 
soundly on a classical {\it stochastical\/} basis. However, the study of 
this latter problem which firstly involves identifying the fundamental 
{\it objects\/}, in this case vacuum oscillators, which are to be 
subjected to the statistical analysis was carried out in terms of 
the equivalent of the one dimensional Schr\"odinger equation. This 
paper is concerned with a more general argument giving a solution 
for the important problem of providing a stochastical foundation for 
Schr\"odinger quantum theory in {\it three\/} dimensions. The step from the 
one dimensional case to the three dimensional case does involve 
some conceptual and technical difficulties. Not the least of which is 
the need to {\it think\/} in six dimensions as this is the least number of 
geometrical dimensions for the configuration space that are 
necessarily invoked in a classical alternative for {\it three\/} 
dimensional quantum theory. Some of the early work on fluid aspects 
of the quantum theory, attempts to reformulation the Schr\"odinger 
structure and discussions of such attempts can be found in 
references [10,20,30,23--26,28]. Work emphasising classical aspects of 
quantum theory can be found in references [2,3]. The statistical 
features of the present work represent a substantial generalisation 
of an idea in reference [21]. Fluid aspects of the present work have 
been discussed in references [17,18,19].

\head 2.   OSCILLATORS \endhead
The six dimensional space in which the alternative theory is 
formulated and in which the {\it physical\/}{\enspace}events underlying the quantum 
process are considered to be taking place arises from the analytic 
continuation into complex planes of the three usual configuration 
coordinates which we shall take to be $ x_1 ,x_2 ,x_3$  in $ E_3^+$, the ordinary 
three dimensional configuration space of common experience. Where 
necessary, the subscripts on configuration  coordinates such as the 
x's will be represented by Greek letters. Thus by making the 
continuations
$$ x_\alpha \rightarrow z_\alpha = x_\alpha+iy_\alpha,\quad \alpha = 1,2,3\eqno{(2.1)}$$
into complex planes in functions originally of the $ x_{\alpha}$ variables only, 
an additional space $E_3^- $ described by the three $ y_{\alpha}$  variables is 
generated. The following work is concerned with showing how events 
in the six dimensional space $ E_6$  produced by the continuation 
process (2.1) directly determine the form taken by the usual 
Schr\"odinger description, how it supports a sound stochastical basis 
for that description and also makes possible explanations not 
available in the orthodox theory. It is assumed that the physics of 
this scheme is taking place in a real $E_6$  space whilst the complex 
features of the Schr\"odinger structure is an aspect of a convenient 
representation as is the case in some {\it fluid\/} contexts [27]. 
However, if we start the theory in a real space as is perfectly 
possible we do eventually have to reform into a complex structure 
to recover the Schr\"odinger equation. Thus here we shall go straight 
to a complex form and note and discuss the possible real 
representation on the way. The discussion will be confined to 
quantum systems with energy states forming a discrete set as this 
type of quantum system is regarded as furthest removed from the 
classical continuum situation. However, systems with a mixed range 
of states both continuous and discrete could be covered by the 
formalism to be discussed with some elaboration. The device that is 
often used in stochastic systems under such circumstances is the 
employment of Stieltjes integrals. In this work, that would require 
some more elaborate discussion of how the energy states are 
distributed in energy space and the summations over the discrete 
state index would need be replaced by Stieltjes integrals over the 
full range of energy states both continuous and discrete.

      In the case of a system with n discrete states of energy $E_j, 
\ j=1\  to\  n$, a complex oscillator for each state $j$ is introduced 
described by a polarisation vector ${\bold q}_j$ associated with a 
{\it mass\/}\ dipole moment $ {\bold q}_jM_j$, where $ M_j$ is the mass equivalent of the 
energy of the state $j,\ E_j = M_jc^2$. The dipole moments are assumed 
formed from positive and negative mass monopoles resulting from 
vacuum polarisation. Such {\it complex\/} oscillators are essentially 
rotating complex units or unitary complex numbers. Thus we set up a 
{\it complex\/} oscillator for each state $j$ with freedom of motion in $ E_3^{+'}$, 
with phase varying with position in configuration space and also 
varying with {\it direction\/} of polarisation in configuration space. The 
prime on $E_3^{+'}$ is to indicate that {\it complex\/} expression in $ E_3^+$ are being 
used as we unfold the structure initially. Such an oscillatory 
process can be described by a complex configuration coordinate 
three-vector ${\bar{\bold q}}_j$,
$${\bar {\bold q}}_j = q_{j,0}\exp (-i\omega_jt){\bold e}_{jp},\eqno{(2.2)}$$
where
$${\bold e}_{jp}= \exp (-i\alpha_j(\bold x,\bold y))\bold e_p\eqno{(2.3)}$$
and
$${\bold e}_p = \Sigma_{\beta=1}^3\exp (-i\gamma_{\beta}){\bold e}_\beta .\eqno{(2.4)}$$
${\bold e}_{jp}$ is a polarisation phase dependent complex direction vector in 
$E_3^{+'}$ including a state dependent and configuration dependent phase 
factor $\exp (-i{\alpha}_j({\bold x},{\bold y}))$ assumed to depend initially on the two 
unspecified vector parameters ${\bold x}$ and ${\bold y}$ in an unrestricted way. That 
is to say that initially any function  of ${\bold x}$ and ${\bold y}$ can be used.
${\bold e}_p$ is a complex direction vector in $E_3^{+'}$ depending on the three constant 
arbitrary phases ${\gamma}_{\beta}$. ${\bold b}_p={\bold e}_p/{\sqrt{3}}$ is a unit vector in
the sense that
$${\bold b}_p^{\phantom{\ast}}\cdot{\bold b}_p^{\ast}= 1\eqno{(2.5)}$$
and consequently,
$${\bold e}_{jp}^{\phantom{\ast}}\cdot{\bold e}_{jp}^{\ast}= 3.\eqno{(2.6)}$$
It is convenient to denote the state dependant part of the phase 
by the {\it real\/} functions  ${1\over 2}{\phi}_j({\bold x},{\bold y},t)$
\linebreak $={\omega}_jt +{\alpha}_j({\bold x},{\bold y})$ so that

$$E_j = \hbar\omega_j = \hbar\partial(\phi_j/{\scriptstyle 2})/\partial t.\eqno{(2.7)}$$

The three space {\it vector\/} components of ${\bar {\bold q}}_j$,
denoted by ${\bar {\bold q}}_{j,\beta}$ then take the form,
$${\bold{\bar q}}_{j,\beta} = q_{j,0}\exp ((\phi_j({\bold x},{\bold y},t)
+2\gamma_{\beta})/2i)\bold e_{\beta}\eqno{(2.8)}$$
with the corresponding complex {\it scalar\/} components being
$${\bar q}_{j,\beta} = q_{j,0}\exp ((\phi_j({\bold x},{\bold y},t)
+2\gamma_{\beta})/2i).\eqno{(2.9)}$$
All the components of ${\bar{\bold q}}_j$ are solutions of the equation of simple 
harmonic motion   ${\partial}^2{\bar{\bold q}}_j/{\partial}t
= -{\omega}_j^2{\bar{\bold q}}_j$ with ${\omega}_j= E_j/\hbar$ . As the structure to 
be unfolded relies totally on vectors such as ${\bar {\bold q}}_j$ this will be seen to 
have the effect of building all Schr\"odinger {\it wave\/} motion on simple 
linear harmonic motion or equivalently planar rotational motion. The 
$E_3^+$ and $E_3^-$ components of another vector ${\bold q}_{j,\beta}$
related to ${\bar {\bold q}}_{j,\beta}$ can now be defined as
$$q_{+j,\beta} = q_{j,0}\cos ((\phi_j({\bold x},{\bold y},t)
+2\gamma_{\beta})/2)\eqno{(2.10)}$$
and
$$q_{-j,\beta} = q_{j,0}\sin ((\phi_j({\bold x},{\bold y},t)
+2\gamma_{\beta})/2).\eqno{(2.11)}$$
We then have $q_{j,\beta}  = q_{+j,\beta}  + iq_{-j,\beta}$
and ${\bar q}_{j,\beta}  = q_{+j,\beta}  - iq_{-j,\beta}$  which is 
why the bar was introduced over the q at the initial stage (2.2). In 
three dimensions, setting $q_{j,0} =2m_0c^2l_0/\sqrt{3}E_j$ gives
a version of the theory with the  {\it physical\/} advantage that monopolar
velocities lie in magnitude between $\pm c$ in both $E_3^+$ and $E_3^-$. This setting of $q_{j,0}$
also represents the assumption that the {\it planar\/} vacuum polarisation
processess are essentially the activation of a fundamental, state independent,
{\it energy\/} dipole moment of {\it magnitude\/} $2m_0c^2l_0/\sqrt 3  = {\hbar}c/\sqrt 3$
for {\it each\/} of the continuation planes with the state involved being determined
by the positive and negative monopolar masses that constitute the activated dipole.

      The  {\it real\/} six dimensional space representation for an 
oscillator in the state $j$ can now be defined as a parameterized 
vector ${\bold q}_j$ given in two sets of three coordinates by,

$${\bold q}_{+j,\beta} = q_{j,0}\cos ((\phi_j({\bold x},{\bold y},t)
+2\gamma_{\beta})/2){\bold e}_\beta\eqno{(2.12)}$$
and
$${\bold q}_{-j,\beta} = q_{j,0}\sin ((\phi_j({\bold x},{\bold y},t)
+2\gamma_{\beta})/2){\bold e}_{\beta}^{\prime}.\eqno{(2.13)}$$
where the real vectors ${\bold e}^{\prime}$ in $E_6$  are equivalent to using a real 
function $f({\bold x})$ to make the continuations and identifications,
$$f({\bold x}){\bold e}_\beta \rightarrow f({\bold x}+i{\bold y})
{\bold e}_\beta\equiv f_1({\bold x},{\bold y}){\bold e}_\beta +f_2({\bold x},{\bold y})
{\bold e}_\beta^{\prime}\eqno{(2.14)}$$
or
$$i{\bold e}_\beta^{\phantom {\prime}}\equiv {\bold e}_\beta^{\prime} .\eqno{(2.15)}$$
It should be emphasised that the phase functions ${\phi}_j$ are strictly 
{\it real\/} functions. This can be regarded as a stability condition on 
the fundamental dipolar oscillators that are being described and 
simply means that these objects, if part of the system, have a 
special permanency in that they do not degrade with the passage of 
time or with being found in displaced positions in space. The vector 
parameters ${\bold x},{\bold y}$ can now be identified as the position
vectors in $E_3^+$ and $E_3^-$ respectively with the phase angles
${\phi}_j({\bold x},{\bold y},t)$ losing some of their arbitrariness.
This results from identifying the ${\bold y}$ as having the components given
by the analytic continuations (2.1) and because the  ${\phi}_j$ are strictly real
they cannot in general be regular functions of the z  and so will be the harmonic
components of such functions. The real {\it six\/} dimensional representations
for the vectors ${\bold q}_j$ and ${\bar{\bold q}}_j$ of the state $j$ are given by,
$${\bold q}_j=(2m_0c^2l_0/{\sqrt 3}E_j){\Sigma}_{\beta=1}^3(\cos ((\phi_j+2{\gamma}_{\beta})/2){\bold e_\beta}
+\sin ((\phi_j+2{\gamma}_{\beta})/2){\bold e}_{\beta}^{\prime}),\eqno{(2.16)}$$
$${\bold {\bar q}}_j=(2m_0c^2l_0/{\sqrt 3}E_j){\Sigma}_{\beta=1}^3(\cos ((\phi_j+2{\gamma}_{\beta})/2){\bold e_\beta}
-\sin ((\phi_j+2{\gamma}_{\beta})/2){\bold e}_{\beta}^{\prime}).\eqno{(2.17)}$$
These are real phase point vectors. They are both indicative 
pointers locating an angular {\it cycling\/} of the system through its 
{\it phase\/} states as clock hands can be used as indicators in relation 
to periodicity in time. ${\bar{\bold q}}$ as given in (2.17) also plays the important 
r\^ole of position of the positive monopole {\it relative\/} to position of 
the negative monopole of the oscillating dipole. Thus it is the polar 
position separation vector for the monopoles of the state $j$. There 
are two important characteristics of these fundamental vacuum 
polarisation units or dipolar oscillators that have been introduced, 
one kinematical and the other dynamical. The kinematic one is that 
the magnitude of the velocities ${\boldsymbol\beta}_j$ and ${\bar{\boldsymbol\beta}}_j$
of the phase vectors on the continuation planes is always $c/{\sqrt 3}$,
as can be seen by differentiating partially with respect to time thus,
$${\boldsymbol{\beta}}_j = {\partial {\bold q}}_j/{\partial t} =
-(c/{\sqrt 3})\Sigma_{\beta=1}^3(\sin ((\phi_j+2\gamma_{\beta})/2)
{\bold e}_{\beta}-\cos ((\phi_j+2\gamma_{\beta})/2){\bold e}_{\beta}^{\prime}),\eqno{(2.18)}$$
$${\bar{\boldsymbol{\beta}}}_j = {\partial {\bar {\bold q}}}_j/{\partial t} =
-(c/{\sqrt 3 })\Sigma_{\beta=1}^3(\sin ((\phi_j+2\gamma_{\beta})/2)
{\bold e}_{\beta}+\cos ((\phi_j+2\gamma_{\beta})/2){\bold e}_{\beta}^{\prime}),\eqno{(2.19)}$$
where (2.7) has been used. It should be emphasised that the circular 
motions described by (2.16--19) is deterministic and takes place at 
the constant speed $c/{\sqrt 3}$ in each of the continuation planes but
that at any moment has the total
speed of light value $c$ in $E_6$. However, the positions on the circles of the phase point
is different in general for the three planes according to the values of the  ${\gamma}_{\beta}$ .
It should also be noted that so far there are no statistical 
considerations involved in the construction. When such statistical 
considerations are involved and consequently there is uncertainty 
of which energy state the system may be in at any moment of time, 
the same uncertainty will apply to the local {\it direction\/} of motion at 
the speed c exhibited by the system. Under those conditions, it is 
appropriate to call the process {\it zitterbewegung\/} [7,8,31]. This later 
process has some resemblance to Brownian motion [20,1,28,25]. The 
important dynamical property of the basic dipolar oscillators 
concerns the angular momentum they induce in relation to the three 
continuation planes. In each of the three continuation planes ${\alpha} =1,2,3$ 
a relative angular momentum ${\bold a}_{ij,\alpha}$ can be defined using the dipole 
moments of the oscillators and the velocities (2.16--19) as follows,
$${\bold a}_{ij,\alpha} = (M_j{\bold q}_{+j,\alpha}-M_i{\bold q}_{-i,\alpha})
{\wedge}({\boldsymbol \beta}_{+j,\alpha}-{\boldsymbol \beta}_{-i,\alpha}),\eqno{(2.20)}$$
with
$${\bold a}_{ji,\alpha} = (M_i{\bold q}_{+i,\alpha}-M_j{\bold q}_{-j,\alpha})
{\wedge}({\boldsymbol \beta}_{+i,\alpha}-{\boldsymbol \beta}_{-j,\alpha}),\eqno{(2.21)}$$
when two different states $i$ and $j$ are employed or as,
$${\bold a}_{jj,\alpha} = (M_j{\bold q}_{+j,\alpha}-M_j{\bold q}_{-j,\alpha})
{\wedge}({\boldsymbol \beta}_{+j,\alpha}-{\boldsymbol \beta}_{-j,\alpha}),\eqno{(2.22)}$$
when the single state $j$ is employed. These are {\it relative\/} angular 
momenta in the sense that they all involve the positive mass 
monopolar field relative to the negative mass monopolar field.

      In evaluating the angular momenta (2.22) from a single dipole 
or (2.20) and (2.21) from interactions between dipoles of different 
energies a difficulty associated with working in six dimensions is 
encountered. In three dimensional space, a {\it pseudo\/} vector product 
such as ${\bold e}\wedge{\bold e}^{\prime}$ has a clean interpretation as
a {\it vector\/} ${\bold k}$ say, orthogonal to the plane of ${\bold e}$ and
${\bold e}^{\prime}$. In six dimensions, any two-dimensional plane has a
four-dimensional space orthogonal to it. Thus a product such as 
${\bold e}\wedge{\bold e}^{\prime}$ does not have a
simple unique vector interpretation in $E_6$. This confronts us with some technical 
difficulties and also with the the possibility that there may be 
advantages in expressing this work in more than the basic six 
dimensions that are being employed here. However, the dimension issue 
will be avoided and left open in the present work by employing a 
self consistent definition for such products that can be used in 
the two contexts that they are needed without giving them a 
physical {\it vector\/} identification. Thus a product such as 
${\bold e}_{\alpha}^{\phantom{\prime}}\wedge{\bold e}_{\alpha}^{\prime}$ will be denoted by
${\bold k}_{\alpha}$ and assumed to have the properties,
$${\bold k}_{\alpha}^{\phantom {\prime}} = {\bold e}_{\alpha}^{\phantom {\prime}}
\wedge{\bold e}_{\alpha}^{\prime}=-{\bold e}_{\alpha}^{\prime}
\wedge{\bold e}_{\alpha}^{\phantom {\prime}},\eqno{(2.23)}$$
$${\bold k}_{\alpha}^{\phantom {\prime}}\cdot{\bold e}_{\alpha}^{\phantom {\prime}}
={\bold k}_{\alpha}^{\phantom {\prime}}\cdot{\bold e}_{\alpha}^{\prime} = 0,\eqno{(2.24})$$
$$\vert{\bold k}_{\alpha}\vert = 1.\eqno{(2.25)}$$
Besides the need to use the ${\bold k}_{\alpha}$ in evaluating the angular momenta 
(2.20--22) they are also necessary for the construction of a curl on 
each of the continuation planes. These curls are then used when 
statistical considerations are introduced in writing down an 
{\it average\/} Maxwell equation for each plane which in turn are used to 
find the electric current densities on the planes. Thus anticipating 
introduction of the statistics, a {\it special\/} $curl\ {\boldsymbol\nabla}_{\alpha}\wedge$
only applicable to {\it vectors\/} that are restricted by the condition of being  
parallel with ${\bold k}_{\alpha}$ is defined by the usual {\it three\/}
dimensional determinantal form,
$${\boldsymbol\nabla}_{\alpha}\wedge({\chi}{\bold k}_{\alpha})=
\left|\matrix
\quad{\bold e}_{\alpha}^{\phantom{\prime}} & \quad{\bold e}_{\alpha}^{\prime} &
\quad{\bold k}_{\alpha} \\
\partial/\partial x_{\alpha} & \partial/\partial y_{\alpha} & 
\partial/\partial{\zeta}_{\alpha} \\
{\phantom {\partial /}}\lower2pt\hbox{0} & {\phantom {\partial /}}
\lower2pt\hbox{0} & {\quad\chi}\endmatrix\right|=
{\bold e}_{\alpha}{\partial}{\chi}/{\partial}y_{\alpha}
-{\bold e}_{\alpha}^{\prime}{\partial}{\chi}/{\partial} x_{\alpha},\eqno{(2.26)}$$
where ${\zeta}_{\alpha}$ is an unspecified extension into the ${\bold k}_{\alpha}$
direction. Thus without identifying the ${\bold k}_{\alpha}$ , (2.26) can be used to
evaluate curls of 
quantities of the form  ${\chi}({\bold x},{\bold y},t){\bold k}_{\alpha}$  which are
like vectors pointing in the direction of ${\bold k}_{\alpha}$  and are related to the
plane ${\alpha}$ . This is a simple way of handling the aforementioned difficulty.
Probably a better, but likely more abstruse scheme can be constructed to deal with this 
type of difficulty.

      The expressions (2.12--13) or (2.16--17) describe the six vector 
parts of a {\it real deterministic\/} oscillatory dipole associated with 
the energy state $j$ located at position $({\bold x},{\bold y})$ in $E_6$ .
It is these expressions that describe the behaviour of what will be regarded as 
a {\it basic\/} type of stable physical unit system. The actual state of 
the vacuum at any place and time will be described by {\it assemblies\/} 
of such basic dipolar oscillators according to the number of 
oscillators for any given energy state $j$ present in the assembly. 
Thus the statistical considerations are introduced by studying 
assemblies of {\it deterministic\/} objects. The physical interpretation 
for (2.12--13) and (2.16--17) is, apart from the increased number of 
degrees of freedom and the additional phases ${\gamma}_{\alpha}$, exactly the same 
as for expressions (2.2) and (2.3) in reference [16]. The positive 
mass monopoles are only free to {\it move\/} in the normal three space 
of common experience and experiment, $E_3^+$, whilst the negative mass 
monopoles are only free to {\it move\/} in the {\it extra\/} space arising from 
analytic continuation, $E_3^-$. Thus ${\bold q}_{+j,\beta}$ in (2.12) is interpreted as the 
extension into $E_3^+$ of a dipole of mass moment $M_j{\bold q}_{+j,\beta}$  whilst
${\bold q}_{-j,\beta}$ is interpreted as the extension into $E_3^-$ of a dipole of
mass moment $-M_j{\bold q}_{-j,\beta}$ .

      In contrast with $E_3$ geometry, it is a feature of the $E_6$  
geometry as used in this system that for any point P in $ E_6$, three 
pairs of vectors such as $({\bold e}_{\beta}^{\phantom{\prime}},
{\bold e}_{\beta}^{\prime}),\ {\beta} =1,2,3$
represent three mutually {\it orthogonal\/} planes intersecting at only the {\it one\/}
point P.\ \ On these planes, the projections of the oscillator motion lie on the 
circular orbits,
$$q_{+j,\beta}^2 + q_{-j,\beta}^2 =(2m_0c^2l_0/{\sqrt 3}E_j)^2, \quad\beta = 1,2,3 \eqno{(2.27)}$$
of radius  $|2m_0c^2l_0 /{\sqrt 3}E_j|$ . However, the phase of these circular 
motions differ from plane to plane according to the values of the 
constant phases ${\gamma}_{\beta},\ \beta =1,2,3$. To analyse the {\it monopolar\/} motions in $E_3^+$   
and $ E_3^-$ it is convenient to introduce new configuration variables $X_{\beta}
=q_{+j,\beta}/q_{j,0}$ for the positive mass monopoles and $Y_{\beta} =q_{-j,\beta}/q_{j,0}$
for the negative mass monopoles. The analysis using these ${\bold X}$ and ${\bold Y}$ 
vectors scales the motions to the same value for all energies so 
that the $j$ subscript is temporarily dropped. We find that for all 
states the motions are confined to two-dimensional planes in $E_3^+$  and 
$E_3^-$ given by,
$$X_1\sin ({\gamma}_3-{\gamma}_2)+X_2\sin ({\gamma}_1-{\gamma}_3)+X_3\sin ({\gamma}_2-{\gamma}_1)
=0\ \eqno{(2.28)}$$
and
$$Y_1\sin ({\gamma}_3-{\gamma}_2)+Y_2\sin ({\gamma}_1-{\gamma}_3)+Y_3\sin ({\gamma}_2-{\gamma}_1)
=0\eqno{(2.29)}$$
respectively. On the plane (2.28), the motion of a positive mass 
monopole lies on an ellipse with projections onto the three 
coordinate planes of $ E_3^+$   given by,
$$X_1^2 +X_2^2 -2X_1X_2\cos ({\gamma }_1-{\gamma}_2) = \sin ^2({\gamma}_1-{\gamma}_2),\eqno{(2.30)}$$
$$X_1^2 +X_3^2 -2X_1X_3\cos ({\gamma }_1-{\gamma}_3) = \sin ^2({\gamma}_1-{\gamma}_3),\eqno{(2.31)}$$
$$X_2^2 +X_3^2 -2X_2X_3\cos ({\gamma }_2-{\gamma}_3) = \sin ^2({\gamma}_2-{\gamma}_3),\eqno{(2.32)}$$
On the plane (2.29), the motion of a negative mass monopole lies on 
an ellipse with projections onto the three coordinate planes of $E_3^-$   
given by,
$$Y_1^2 +Y_2^2 -2Y_1Y_2\cos ({\gamma }_1-{\gamma}_2) = \sin ^2({\gamma}_1-{\gamma}_2),\eqno{(2.33)}$$
$$Y_1^2 +Y_3^2 -2Y_1Y_3\cos ({\gamma }_1-{\gamma}_3) = \sin ^2({\gamma}_1-{\gamma}_3),\eqno{(2.34)}$$
$$Y_2^2 +Y_3^2 -2Y_2Y_3\cos ({\gamma }_2-{\gamma}_3) = \sin ^2({\gamma}_2-{\gamma}_3),\eqno{(2.35)}$$
In the equations for these orbits, the differences relating to 
different energies have been scaled away by introducing the new 
variables ${\bold X}$ and ${\bold Y}$. However, it is clear from the scale factor 
$2m_0c^2l_0/{\sqrt 3}E_j$  that the larger energy orbits are nested {\it within\/} the 
smaller energy orbits.

      The projections of the motions onto the three continuation 
planes spanned by the pairs of vectors $({\bold e}_{\beta}^{\phantom{\prime}},
{\bold e}_{\beta}^{\prime}),
\ {\beta}  =1,2,3$ given by the 
circles (2.27) play a very important part in the development of the 
theory as these motions generate the {\it internal\/} angular momentum 
(2.20--22) additional to any angular momentum arising from the motion 
(2.30--35) that might be present in $E_3^+$ or indeed present in $E_3^-$. This 
internal angular momentum arises in discrete units of the size of 
one third of one Planck's constant on each of the three continuation planes 
for every {\it single\/} oscillator that might be present at the location 
$({\bold x},{\bold y})$ as can be seen by using the vector products (2.23--25) to 
evaluate the angular momenta ${\bold a}_{ij}$ defined at (2.20--22). We find that 
the interaction angular momentum for two distinct oscillators in 
energy states $i$ and $j$ involves the two contributions given by
$${\bold a}_{ij,\beta} = {\bold a}_{ji,\beta} = -(2m_0cl_0/3)\cos (({\phi}_i-{\phi}_j)/2)
{\bold k}_{\beta} \eqno{(2.36)}$$
while the contribution from one oscillator in {\it self\/} interaction is,
$${\bold a}_{ij,\beta} = -(2m_0cl_0/3){\bold k}_{\beta} =
-(\hbar/3){\bold k}_{\beta},\eqno{(2.37)}$$
with the last quantity having the magnitude of one third of one Planck unit ${\hbar}$ . 
(2.36) and (2.37) both hold for the three continuation planes ${\beta} =1,2,3$. 
Thus the basic {\it objects\/} with their characteristics about which in 
{\it assembled\/} form a statistical or stochastical analysis is to be 
conducted are identified.  These basic objects are dipoles though 
not of the limiting form of vanishing length and infinite pole 
strength sometimes employed in other contexts but rather dipoles 
with finite separation of finite mass weighted pole and antipole. The 
positive mass pole in motion on one of the ellipses described above 
in normal configuration space with the negative mass antipole in 
motion on the corresponding ellipse in the three space generated by 
analytic continuation. The relative angular motion of the  
constituent monopolar pair generates an angular momentum in units 
of ${\hbar/3}$ as three  {\it vectors\/} orthogonal to the projected motions on 
the three continuation planes. Any one such dipole has a definite 
energy associated with it given by the formula (2.7), where 
${\phi}_j({\bold x},{\bold y},t)/2$  is the phase angle of the oscillating dipole. The 
restrictions on how the functions ${\phi}_j/2$  depend on the variables  
${\bold x},{\bold y}$ and $t$, are, so far, firstly the condition (2.7) relating
them to the energy $E_j$ involving time and secondly the condition that they should 
be harmonic if the ${\bold y}$ parameter is a vector with components 
identified with the imaginary parts of the analytic continuation 
variables. However, it should be remarked that the class of harmonic 
functions is a very large class of functions so that the harmonic 
condition is not very restrictive. Clearly, dipolar objects such as 
we have defined via their energy and kinematic properties can be 
conceived as arising out of the {\it zeroness\/} of the vacuum by 
{\it polarisation\/}. The nature and properties and statistical aspects of 
assemblies of such objects spread about configuration space and 
their relation with three dimensional Schr\"odinger theory will be 
discussed in the next sections.

             \head 3  ASSEMBLIES \endhead

      Having identified the oscillatory dipolar basic objects, it is 
now possible to consider assemblies of them involving collections 
locally with numbers of individual members at various energies $E_j$ 
and with different phases at different positions in $E_6$ . This 
approach represents an uncomplicated, unambiguous and indeed a 
standard way to introduce probabilistic considerations into the 
analysis of an initially deterministic situation. Firstly, let us 
consider a local collection of oscillators composed of $n_i$ members in 
the energy state $E_i\  for\quad i=1\  to\  n$. The total internal angular 
momentum on the plane ${\beta}$  generated by all the $n^2$  interactions in the 
assembly between the oppositely signed fields from different energy 
oscillators and including self interactions is
$${\bold a}_{\beta}={\Sigma}_{ij}n_in_j{\bold a}_{ij,\beta} = -(\hbar /3){\bold k}_{\beta}(
{\Sigma}_in_i\exp ({\phi}_i/2i))({\Sigma}_in_i\exp (-{\phi}_i/2i)),\eqno{(3.1)}$$
where (2.36--37) have been used. Denoting the factors in (3.1) by ${\Psi}$  
and  ${\Psi}^{\ast}$ which are in this paper dimensionless functions gives,
$${\Psi}({\bold x},{\bold y},t) = {\Sigma}_in_i\exp (+{\phi}_i({\bold x},{\bold y},t)/2i),\eqno{(3.2)}$$
$${\Psi}^{\ast}({\bold x},{\bold y},t) =
{\Sigma}_in_i\exp (-{\phi}_i({\bold x},{\bold y},t)/2i),\eqno{(3.3)}$$
and
$${\bold a}_{\beta} = -(\hbar /3){\bold k}_{\beta}{\Psi}^{\ast}({\bold x},{\bold y},t)
{\Psi}({\bold x},{\bold y},t) =-(\hbar /3){\bold k}_{\beta}
{\rho}^{(0)}({\bold x},{\bold y},t),\eqno{(3.4)}$$
having introduced the useful dimensionless function ${\rho}^{(0)} ={\Psi}^{\ast}{\Psi}$.
The dependence of the functions ${\Psi}^{\ast}$ and ${\Psi}$ on ${\bold x}$ and
${\bold y}$ only arising through the harmonic functions  ${\phi}_i{(\bold x},{\bold y},t)$
involved in their structure. It is interesting to consider the total {\it internal\/}
angular momentum ${\bold a}$ contributed by the motions on all the continuation planes,
Using (3.4), this can be written in the form $\bold a =-(\hbar {\bold k}/\sqrt 2)N_0$,
where the unit vector ${\bold k} =
{\Sigma}_{\beta =1}^3 (1/\sqrt 3){\bold k}_{\beta}$ has been introduced
and $N_0=\sqrt{2/3}\rho^{(0)}$.
The unit vector ${\bold k}$
can have  components in $E_3^+$ and $E_3^-$. Denoting these components by $\cos (\chi )$
and $\sin (\chi )$ respectively we see that ${\bold a }$ can be written in the form,
${\bold a ={-(\hbar}/\sqrt 2)}({\bold k}_+\cos (\chi)-{\bold k}_-\sin (\chi))N_0$.
Confining our attention to the one special case when the angle $\chi$ has the value $\pi /4$,
the total relative internal angular momentum
becomes ${\bold a}= -(\hbar /2)({\bold k}_+-{\bold k}_-)N_0$. Thus in this case the
total relative internal angular momentum ${\bold a}$ is composed of $N_0$ differences of half
integral {\it spin \/} units. If the spins in the $E_3^+$ direction and the $E_3^-$
direction are regarded as coming from the positive and negative monopoles respectively
then it becomes apparent that the polarisation process under consideration in this paper
is responsible for the generation of the equal {\it magnitude\/} spin vectors
$(\hbar /2){\bold k}_+$ and $(\hbar /2){\bold k}_-$. However, it should be noted that being in orthogonal spaces,
the {\it directions\/} of these spins are not opposite in the way that opposite monopolar
charges or masses are opposite. The other three special cases given by
$\chi = 3\pi /4, 5\pi /4$ and $7\pi /4$ which are related to spin flip through an angle of $\pi /2$
or spin reversal can be analysed in a similar manner to the $\pi /4$ case. 
Thus spin and vacuum polarisation appear to be very closely related
and the vacuum polarisation structure used in this theory implies the existence of spin.

The derivation given in this paper, of the form (3.4) from the energy state 
assembly contains the solution to the long standing conceptual 
problem of why, when in quantum mechanics state functions are 
superposed as in (3.2) and (3.3), probability takes on a bilinear form 
such as is contained in (3.4). Such a structure is not found in what 
is called classical probability theory. It emerged in quantum theory 
in the form of an ad hoc rule for constructing probabilities from 
wave functions.
It is clear that the bilinear form $N_0 =\sqrt{2/3}{\rho}^{(0)}$
gives the number of Planck {\it half\/} angular momentum units arising from all of the three
continuation planes generated by all the interactions between dipolar oscillators in
the {\it assembly\/} of states of varying energy contributed by the positive mass monopoles.
It similarly gives the number of subtracted half spin units contributed by the negative mass monopoles.
Thus the number of {\it Planck\/} units depends 
directly on the {\it higher\/} assembly of {\it interactions\/} rather than as 
one might at first expect on the assembly of energy states. In this 
version of the quantum structure, the r\^oles of and relations 
between these two assemblies is totally transparent. However, in 
the orthodox theory there is no explanation for the relation 
between the bilinear form and the superposition principle other than 
the ad hoc rule of thumb for constructing probabilities from wave 
functions.

      The quantity  $N_0$ is essentially the {\it number\/} of Planck half units
contributed by monopoles activated by the planar motions and generated by 
the angular momentum interactions between the energy states. It is 
also effectively the number measure of various other physical 
quantities induced or present in the $E_6$  space. Assuming that such
quantities are produced in or reside in some standard constant {\it volume\/}
${v_0}$ which according to context represents six, three or the two dimensional
volume we call area, then a quantity ${\rho}  = {N_0}/v_0$\ can be 
defined that will apply throughout the space and represent number 
{\it density\/} for whatever dynamical quantity is under discussion be it 
angular momenta units, electric dipoles, magnetic dipoles or indeed 
some other local characteristic quantity. Such a number density is, 
of course, not a probability density but often it can be converted 
to a probability density by suitable normalisaton. Thus if it is 
desired to keep strictly within the remit of probabilistic stochastic 
theory the number frequencies $n_j$ for energy states present in an 
assembly and the interaction number frequencies $n_in_j$ for 
interactions between energy states can be used to define the 
{\it two-level\/} linear and quadratic sets of probabilities $P_j$ and $P_{ij}$,
$$P_j = n_j/N,\quad N = {\Sigma}_in_i\eqno{(3.5)}$$ 
and
$$P_{ij} = n_in_j/N_f, \quad N_f = ({\Sigma}_in_i)^2.\eqno{(3.6)}$$
The linear probabilities $P_j$ are directly related to the linear 
superposition principle of quantum mechanics that is generally used 
to combine eigenstates as in (3.2). The {\it quadratic\/} probabilities $P_{ij}$ 
are closely related to the probabilities usually employed in quantum 
mechanics and they measure the weight carried by the interactions 
between different states and include self interactions of single 
states as in the angular momentum structure discussed earlier. Thus 
two different {\it expectation\/} values can be defined; a linear one $E_L[\,.\,]$ 
for use with averaging vector like quantities such as a set of 
unitary exponential oscillator amplitudes ${\bold {\eufm u}}=\{\exp i{\phi}_i\}$ and a
quadratic one $E_Q[\,.\,]$ for use with averaging matrix like quantities such as the 
set of angular momentum interactions between vector like components 
${\bold {\eufm a}}=\{{\bold a}_{ij}\}$. Thus using the probabilistic formulation
the wave function can be expressed as,
$${\Psi} = NE_L[{\eufm u}],\eqno{(3.7)}$$
while the number function ${\rho}^{(0)}$ can be expressed as,
$${\rho}^{(0)} = - {\sqrt 3}N_fE_Q[{\eufm a}\cdot{\bold k}/{\hbar}].\eqno{(3.8)}$$
The clear separation of these different statistical measures should 
prevent the intrusion and the assuming of undue importance by 
those {\it paradoxical\/} issues [23,29,5] that arise from confusing the 
linear and the quadratic assembly. The connection between the linear 
superposition principle and the number of Planck units to be found 
at any position on the configuration planes is an aspect of a wider 
issue of the significance of the nonlinear exponential 
transformation that the wave function represents. The insight used 
here with regard to how the number of Planck units depends on the 
{\it higher\/} assembly of interactions and its significance for quantum 
theory, was a direct result of a conversation between the Author 
and Professor M. Atiyah on nonlinearity in 1986.

      It has been implicitly assumed so far that given the harmonic 
functions ${\phi}_j({\bold x},{\bold y},t)$, the assembly of energy states
is constructed with the two vector parameters ${\bold x},{\bold y}$ fixed.
When these two parameters 
are taken to represent a position in $E_6$ this implicit assumption is 
equivalent to setting up the energy assembly at the six-point $({\bold x},{\bold y})$. 
The space dependence of the assembly can be further extended by 
allowing the assembly frequencies $n_j$ to themselves depend on 
position. This has the effect of setting up differing energy 
assemblies from point to point in $ E_6$ . Taking this step, we now 
introduce a set of {\it arbitrary\/} assembly frequencies $n_i$ as follows,
$$n_j = N_jn_{cj}({\bold x},{\bold y}), \quad j=1\  to\  n.\eqno{(3.9)}$$
The functions $N_j$ are absolute constants and the function $n_{ci}({\bold x},{\bold y})$ 
depend only on ${\bold x}$ and ${\bold y}$. The function ${\Psi}$  introduced at (3.2) now 
assumes the form,
$${\Psi}({\bold x},{\bold y},t) = {\Sigma}_iN_in_{ci}({\bold x},{\bold y})
\exp ({\phi}_i({\bold x},{\bold y},t)/2i).\eqno{(3.10)}$$
If the ${\bold y}$ vector in these functions is identified as arising from the 
analytic continuation process (2.1) or  ${\Psi}({\bold x},{\bold y},t)$ is taken
to be of the form  ${\bold {\Psi}}({\bold x}+i{\bold y},t)$, it then follows that
$\ln (\Psi ({\bold x},{\bold y},t))$ can be taken to be a 
regular function of the complex vector ${\bold z} = {\bold x}+i{\bold y}$.
Uniformity of structure and the fact that the ${\phi}_i$ are harmonic then implies
that the same argument can be applied to every term in the finite sum in (3.10) to 
give the result that the {\it eigen\/} complex fluid potential functions,
$$ w_i(z) = 2{\nu}({\phi}_i({\bold x},{\bold y},t)/2
+i\ln (n_{ci}({\bold x},{\bold y})),\eqno{(3.11)}$$
are regular functions of ${\bold z}$. This is a dramatic conclusion because it 
implies that although the basic oscillators are stable against 
degradation in time or space their phase is {\it soft\/} in the sense 
that their distribution {\it information\/} in space as given by $\ln (n_{ci}({\bold x},{\bold y}))$ 
effects their phase angles as given by ${\phi}_i({\bold x},{\bold y},t)/2$ because these two 
function are related by harmonic conjugacy. Thus it is impossible to 
set up an {\it arbitrary\/} space distribution of assembled energy 
oscillators with preset harmonic phases in the structure being 
examined here. The conclusion is {\it dramatic\/} because it has been 
arrived at without any use of a Schr\"odinger or equivalent equation 
that would mould the spatial form of dependence of the ensembles. 
It leads to the significant conjecture that at some fundamental level 
information and phase are linked each being obtainable from the 
other by methods such as the Hilbert transform. This, of course, is 
reminiscent of the {\it causality\/} context and its attendant dispersion 
relations [32].

                 \head 4 CURRENTS\endhead

      The angular momenta (3.4) arising from the oscillators and 
associated with the continuation planes is a consequence of a 
relative rotational motion of the state monopoles as given by the 
circular orbits (2.27). As the state monopoles carry charges 
proportional to their masses it is to be expected that the same 
rotational motion will produce a magnetic moment which can be 
obtained from the angular momentum ${\bold a}$  using the {\it classical\/} 
gyromagnetic ratio, $-|e| /2m_0$. This gives for the magnetic moment 
{\it density\/} induced on the continuation planes,
$${\bold g}_{mag,\beta} = +(|e|cl_0/3){\bold k}_{\beta}{\rho}^{(0)}({\bold x},{\bold y},t)/v_0
=(|e|l_0c/\sqrt 6){\bold k}_{\beta}\rho ({\bold x},{\bold y},t),\eqno{(4.1)}$$
where the dipole density function ${\rho}$  has been introduced via a 
standard constant comparison volume $v_0$ as $\rho = N_0/v_0$ .
The equivalent magnetic induction vector for the plane can then 
be written as,
$${\bold B}_{0,\beta} = {\mu}_0{\bold g}_{mag,\beta}.\eqno{(4.2)}$$
This quantity is a characteristic of the {\it assembly\/} depending as it 
does on the frequencies $n_j$ of the distribution of energy states. If 
${\bold B}_{0,\alpha}$ is used in a Maxwell equation on the plane ${\alpha}$ to determine the 
surface electric currents that it may be regarded as having induced 
or that it may have been induced by, then these currents 
themselves will be characteristic of the assembly. That is to say the  currents.
are in some sense {\it average\/} currents. The three Maxwell equations to be used are,
$${\mu}_0{\bold J}_{\alpha} = {\boldsymbol\nabla}\wedge ({\bold B}_{0,\alpha})
,\quad \alpha = 1,2,3.\eqno{(4.3)}$$
The ${\bold J}_{\alpha}$  are {\it three\/} two-dimensional current densities for the three 
planes. We see from (4.1) that the densitry of Bohr magneton units $|e|cl_0$ generated
on the continuation planes is $\rho /{\sqrt 6}$. Thus for consistency we use a Maxwell equation
to relate this to a current density ${\bold J}_{\alpha}$ with the numerically same density
of electric charges $|e|$. This is where the special curls (2.26) are needed and when 
used in (4.3) with (4.2) the currents ${\bold J}_{\alpha}$  are found to be
$${\bold J}_{\alpha} = -(|e|/{\sqrt 6}){\rho}({\bold x},{\bold y},t)
{\bar{\boldsymbol \beta}}_{\alpha},\eqno{(4.4)}$$
where
$${\bar{\boldsymbol\beta}}_{\alpha} = {\bold u}_{\alpha} -{\bold v}_{\alpha}\eqno{(4.5)}$$
or in terms of ${\rho}^{(0)}$
$${\bar{\boldsymbol\beta}}_{\alpha} = -cl_0({\partial}ln{\rho}^{(0)}({\bold x}.{\bold y},t)/
{\partial}y_{\alpha}{\bold e}_{\alpha}-{\partial}ln{\rho}^{(0)}({\bold x},{\bold y},t)/
{\partial}x_{\alpha}{\bold e}_{\alpha}^{\prime}).\eqno{(4.6)}$$
${\bar {\boldsymbol\beta}}$ given by (4.5) is an {\it average\/}
local velocity field of the positive 
monopoles relative to the negative monopoles on the plane  . So 
far, the the only basic concepts introduced are vacuum polarisation, 
analytic continuation and assemblies with their statistical 
connotations. However, the structure that has emerged looks much 
like the usual Schr\"odinger theory. In fact, a generalisation of 
orthodox three-dimensional linear Schr\"odinger theory has been 
produced. All that is needed to recover the subcases of linear or 
indeed nonlinear Schr\"odinger theory is the important {\it classical\/} 
constraint of some type of continuity equation for density 
describing its transport about the configuration space and possible 
input or output by external interference. It turns out that the 
continuity equation is the source of both the Schr\"odinger equation 
for the system and also of the space time governing stochastic 
differential equation. Equivalent equations of flow continuity 
can be written down in $E_3^+$, $E_3^-$  or in $E_6$.
In the latter case, the appropriate form is,
$$\partial\rho /\partial t = -{\boldsymbol\nabla}\cdot (\rho{\boldsymbol\beta}^{(c)})
 + \Gamma +{\Gamma}_{fb},\eqno{(4.7)}$$
where ${\boldsymbol\nabla}\cdot$ is the six-dimensional divergence in $E_6$
and ${\boldsymbol\beta}^{(c)}$ is the six 
dimensional centroidal monopolar velocity field  ${\boldsymbol\beta^{(c)}
=(\bold u} + {\bold v})/2$ for 
monopolar pairs and is obtainable from the ${\bold u}$, ${\bold v}$ components of (4.5) 
or (4.6). The source term ${\Gamma}$ which is of the form
${\rho}V_2({\bold x},{\bold y})/m_0cl_0$ is 
{\it always\/} necessary if there is an external field present but is 
generally zero on the subspace $E_3^+$  where ${\bold y}=0$ as $V_2$ is the imaginary 
part of the {\it usually\/} real external potential $V_1({\bold x})$. That is to say 
$V_2({\bold x},{\bold y})$ is chosen to be the harmonic conjugate of some specific 
function $V_1({\bold x},{\bold y})$ so that together they form a regular function 
$V({\bold z})=V_1({\bold x},{\bold y})+iV_2({\bold x},{\bold y})$ of ${\bold z}$
=${\bold x}$+i${\bold y}$. This choice of $V_2$ is important for the 
derivation of the Schr\"odinger equation when $ V_1$ is recognised as the 
usual external potential. The source term ${\Gamma}_{fb}({\bold x},{\bold y},t)$
will be taken to be zero in this paper but is an additional {\it feedback\/} contribution 
involving some functional form of the density  that is necessary if 
the structure is used to analyse nonlinear Schr\"odinger systems [17]. 
If the expressions for ${\bold u}$ and ${\bold v}$ are substituted into (4.7), the 
equation of centrifugal diffusion in $E_6$ ,
$${\partial}{\rho}/\partial t = \nu {\Sigma}_{\alpha=1}^3{\partial}^2{\rho}
/\partial x_{\alpha}\partial y_{\alpha} + \Gamma ,\eqno{(4.8)}$$
is obtained with centrifugal diffusion constant ${\nu} =cl_0$ . The feedback 
term is omitted from (4.8) onwards while we concentrate on linear 
Schr\"odinger theory. Equation (4.8) is the basic {\it stochastical\/} 
differential equation that governs the process described by the 
distribution   as it unfolds in the six dimensional configuration 
space against time. This equation fully explains the structure of 
orthodox {\it linear\/} Schr\"odinger theory as a {\it rotatory\/} diffusive 
{\it dynamical\/} equilibrium of the dipolar density ${\rho}$ in $E_6$ [19].

      The Schr\"odinger equation is obtained from (4.8) by 
substituting into it the bilinear form for the density
${\rho} ={\sqrt{2/3}} {\rho}^{(0)}({\bold x},{\bold y},t)/v_0$  
formed from ${\Psi}^{\ast}$ and ${\Psi}$ in (3.4).
The quickest route to the Schr\"odinger 
equation is to replace the second order differential operators in 
(4.8) by their ${\bold z},{\bar{\bold z}}$ equivalents,
$${\partial}^2/\partial x_{\alpha}\partial y_{\alpha}
 = i({\partial}^2/\partial z_{\alpha}^2 - {\partial}^2/\partial {\bar z}_{\alpha}^2) \eqno{(4.9)}$$
with
$${\rho}^{(0)}({\bold x},{\bold y},t) = {\Psi}^{\ast}({\bar{\bold z}},t)
{\Psi}({\bold z},t).\eqno{(4.10)}$$
The {\it asterisk\/} superscript on ${\Psi}$ denotes the complex conjugate of the
functional {\it form\/} only. Thus the full complex conjugate denoted by 
the top {\it bar\/} is given by ${\bar{\Psi}}({\bold z},t) = {\Psi}^{\ast}({\bar{\bold z}},t)$.
We find that
$$V_2({\bold x},{\bold y}) = 2Im.((m_0\nu/{\Psi}({\bold z},t))(i{\partial}{\Psi}({\bold z},t)
/{\partial}t+{\nu}{\Sigma}_{\alpha = 1}^3{\partial}^2{\Psi}({\bold z},t)
/{\partial}z_{\alpha}^2)).\eqno{(4.11)}$$
However, $V_2({\bold x},{\bold y})$ was chosen to be the {\it harmonic\/} conjugate of some 
definite function $V_1({\bold x},{\bold y})$. It follows that there is a {\it regular\/} 
function $V({\bold z})$ of the complex vector ${\bold z}$ such that
$$V(\bold z) = V_1({\bold x},{\bold y}) + iV_2({\bold x},{\bold y})
=(\hbar /{\Psi}({\bold z},t))(i{\partial{\Psi}}({\bold z},t)/{\partial}t
+ {\hbar}/2m_0{\Sigma}_{\alpha = 1}^3
{\partial}^2{\Psi}({\bold z},t)/{\partial}z_{\alpha}^2).\eqno{(4.12)}$$
From (4.12), the analytically continued three dimensional linear 
Schr\"odinger equation with complex external potential $V({\bold z})$ follows,
$$i{\hbar}{\partial}{\Psi}({\bold z},t)/{\partial}t = -({\hbar}^2/2m_0)
{\Sigma}_{\alpha = 1}^3{\partial}^2{\Psi}({\bold z},t)/{\partial}z_{\alpha}^2
+V(\bold z){\Psi}({\bold z},t)\eqno{(4.13)}$$
and this clearly reduces to the usual $E_3^+$ Schr\"odinger equation,
$$i{\hbar}{\partial}{\Psi}({\bold x},t)/{\partial}t = -({\hbar}^2/2m_0)
{\boldsymbol\nabla}_{\bold x}^2{\Psi}({\bold x},t)
 +V_1({\bold x){\Psi}({\bold x},t}).\eqno{(4.14)}$$
of the orthodox theory on the boundary ${\bold y}=0$, where usually $V_2$ =0.

                 \head 5 CONCLUSIONS \endhead

      In this paper, it has been shown that {\it three\/} dimensional 
Schr\"odinger quantum theory can be soundly based on classical 
stochastic theory by setting up a {\it simple\/} probabilistic structure 
using assemblies of oscillating mass dipoles structured according 
the frequencies of occurrence of discrete dipolar energy values.  
Classical physical ideas {\it only\/} are employed in this generalisation 
and reformulation of quantum theory within a framework of not 
greatly advanced or abstruse mathematics. The term {\it classical\/} is 
emphasised here and as far as the physics is concerned it means 
essentially the type of system and formalism that is presented in 
books on dynamics, hydrodynamics and electro-fluids such as 
contained in references [22], [9] and [27] where {\it operators\/} are not 
used to represent dynamical quantities. The classical stochastic 
theory can be found in reference [1] and is characterised by the 
involvement of firmly established  probability theory or standard 
statistical method without involving any {\it new\/} forms of {\it logic\/} 
introduced to patch up paradoxes [29] real or apparent. With regard 
to the mathematics employed in this new foundation for quantum 
theory, it has been the author's policy to use only straight forward 
or {\it hesitatingly\/} what one might call {\it simple\/} mathematical methods 
such as is used in the very clear treatment of complex variable 
theory that can be found in reference [4]. However, it is likely 
that the formalism might well be {\it better\/}  expressed and further 
advanced conceptually by the use of higher level techniques such as
differential forms [6].

In this work, heavy use is made of the idea of vacuum 
{\it polarisation\/} by employing the negative energy states that first 
assumed importance in {\it classical\/} relativity theory when Einstein 
introduced the energy equation $(E/c)^2 =p^2 +(m c)^2$  for the free 
relativistic particle with its two energy states $E= \pm |E|$ . However, 
{\it vacuum\/} polarisation has assumed great importance even in the 
orthodox theory in the development of the theory of quantized 
fields [31] where it is deeply involved in the r\^ole of enabling the 
{\it divergent\/} mathematics of that area to be patched up by 
{\it renormalisation\/} so as to yield finite measurable information. Thus 
vacuum polarisation has to be used to rescue orthodox theory from 
very severe internal mathematical difficulties. Dirac [11] made a 
drastic contribution to the topic of vacuum polarisation when he 
introduced the postulate that the sea of negative energy states 
were {\it occupied\/}. This was to eliminate difficulties with the negative 
energy concept such as state evolution in reversed time and it led 
eventually to the real particle, the anti-electron, or positron being 
identified in the orthodox theory.  However, it had the effect of 
drenching the negative energy states almost out of existence except 
that their contribution had to be taken into account in the form of 
effectively {\it minute\/} corrections to measurable quantities calculated 
by the renormalisation process introduced to render quantum 
electrodynamics a viable theory. Certainly at the technical level 
Dirac's postulate made possible great progress with the structural 
edifice of quantized fields. From the point of view of obtaining 
philosophical understanding of the basis of quantum theory at the 
analytic level, Dirac's postulate disposed of the baby with the bath 
water. In the scheme described in the present paper, the negative 
energy states are dealt with in a way that is not {\it entirely\/} 
different from that suggested by Dirac. Here negative energy states 
are used in a {\it balanced\/} way with the positive energy states so that 
together they form the statistical and dynamical net of the 
{\it quadratic assembly\/} and consequently, while in a sense mutually 
cancelling as positive and negative monopoles, they together cradle 
the actual measurable energy of the system. An Additional interesting 
result that is produced by this work is the {\it classical\/}
explanation for {\it spin\/} as being a type of {\it vector monopolar
unit\/} attached to the basic mass monopoles and generated by vacuum polarisation. This connection has emerged as an unexpected bonus from this hyperspace alternative to Schr\"odinger theory. The work seems to imply that spin as a kinematic {\it rotational\/} polarisation effect is at the foundation of {\it all\/} quantum theory.

Confirmation that this stochastically founded structure is a {\it generalisation\/} of Schr\"odinger theory has now been demonstrated [33] by its application to obtain the {\it very\/} simple formula for the fine structure constant, $\alpha =\cos (\pi /N)/N$.
This formula gives the value of $\alpha$ to very great accuracy when $N$ is set equal to $137$.
$\alpha =\cos (\pi/137)/137\approx 7.297351\  10^{-3}$.
The numerical value given by this formula has an error in relation to the latest
experimentally determined value [34] that is of the order of $2$ parts in $10^9$. 

It has been shown that the relativistic negative energy 
states can be used to form a completely sound stochastic basis for 
three dimensional Schr\"odinger theory with the form of that basis 
having a high geometrical and structural {\it visualisability\/} in terms 
of classical assemblies of orbiting  monopolar particles. This is in 
striking contrast to the {\it deadpan\/} visage of the admittedly 
otherwise technically highly efficient orthodox theory. The author 
believes that the accessibility to the underlying quantum process 
that the present theory achieves, both in terms of structure 
recognition and picturability of the processess taking place in the 
quantum vacuum background, will lead to progress with quantum 
fundamentals and also lead to a better understanding of space time 
structure in general. Certainly, the work puts the quantum process 
into a clearer perspective with regard to its relation to precursor 
theories, showing it to be of the same family and not an alien 
mutation with its own logic. One of the important conclusion that 
stands out from the present work is that although the orthodox 
theory does involve physical jumps and discontinuities in energy and 
other physical variables the extrapolation to the idea that quantum 
mechanics is a {\it philosophical\/} jump away from the older classical 
ideas is false. This holds the implication that if there are 
paradoxes [29,5] present in the quantum theory that are {\it real\/} 
rather than apparent, the same paradoxes must already be present 
in the classical ways of thinking about the physical world. 
\vskip 1cm
\noindent ACKNOWLEDGEMENTS: I am very grateful to Professor M.S. Bartlett for 
the remarks that many years ago set me off on the search for a 
sound stochastic basis for quantum theory and to Professor Sir 
Michael Atiyah who more recently drew my attention to the central 
importance of the nonlinear relation between the quantum fluid 
complex potential and the space of wave function linear 
superposition.


\vskip 1cm
\Refs
\refstyle{A}
\widestnumber\key{34}
\ref\key 1
\by  Bartlett, M.S.\book An introduction to Stochastic Processes
\publ Cambridge University Press \yr 1966
\endref
\ref\key 2
\by Barut, A.O.\paper Class. Quantum Gravity \vol 4 \pages 141 \yr1987
\endref
\ref\key 3 \by Barut, A.O.\publaddr International Centre for Theor. Phys.
\paper Preprint No. 157, Triest \yr1987\endref
\ref\key 4\by Beardon, A.F.\book Complex Analysis\publ John Wiley and Sons Ltd \yr1979\endref
\ref\key 5\by Bohm, D., Hiley, B.J., Kaloyerou, P.N.\paper An Ontological
Basis For The Quantum Theory\jour Physics Reports \vol144\issue 6\pages 323-349
\publaddr North-Holland, Amsterdam \yr 1987\endref
\ref\key 6 \by Cartan, H. \book Differential Forms\publ Kershaw Publishing Co. Ltd. \yr1971\endref
\ref\key 7 \by  Cavalleri, G.\jour LETTERE AL NUOVO CIMENTO \vol43\issue 6 16 Luglio \yr1985\endref
\ref\key 8 \by Cavalleri, G., Spavieri G.\jour IL NUOVO CIMENTO \vol. 95B\issue 211 Ottobre \yr1986\endref
\ref\key 9 \by Cowing, T.G.\book Magnetohydrodynamics\publ Adam Hilger \yr1976\endref
\ref\key 10 \by De Broglie, L.\book La Thermodynamique de la Particle
Isol\'ee \publ Gauthier-Villars\publaddr Paris \yr1964\endref
\ref\key 11 \by Dirac, P.M.\book The Principles of Quantum Mechanics\publ Oxford University Press
\yr 1947\endref
\ref\key 12 \by Gilson, J.G.\paper A classical basis for quantum mechanics
\publ Ann. Inst. Henri Poincare \vol XXX11\issue 4\pages 319--325\yr 1980\endref
\ref\key 13 \by Gilson, J.G.\paper The nature of a quantum state\publ Acta Physica Hungarica
\vol 60\issue 3-4 \pages 145--160 \yr 1986\endref
\ref\key 14 \by Gilson, J.G.\paper An alternative view of the Schr\"odinger
Process\publ Speculations in Science and Technology\vol 9\yr 1986\endref
\ref\key 15 \by Gilson, J.G.\paper Six-dimensional flow and the three
dimensional Schr\"odinger equation\jour Journal of Mathematical
and Physical Sciences\vol 20\issue 1,feb.\pages 55--63 \yr1986\endref
\ref\key 16\by Gilson, J.G.\paper Oscillations of a Polarizable Vacuum
\jour Journal of Applied Mathematics and Stochastic Analysis
\vol 4\issue 11, summer \page 95--110\yr 1991\endref
\ref\key 17\by Gilson, J.G.\paper Classical Fluid Aspects of Nonlinear
Schr\"odinger Equations and Solitons Journal of Applied
Mathematics and Simulation \vol 1\issue 2\pages 99--114\yr 1987\endref
\ref\key 18\by Gilson, J.G.\paper Classical Variational Basis for Schr\"odinger
Quantum Theory\jour Journal of Applied Mathematics and Simulation
\vol 1\issue 4\pages 287-303\yr 1988\endref
\ref\key 19\by Gilson, J.G.\paper Centrifugal Diffusion Negative Mass and the
Schr\"odinger Equation\jour Math. Scientist\vol 5\pages 79--89\yr1980\endref
\ref\key 20\by Gilson, J.G.\paper On Stochastic Theories of Quantum Mechanics
\jour Proc. Camb. Phil. Soc.\vol 64\pages 1061--1070 \yr1968\endref
\ref\key 21\by Gilson, J.G.\paper Statistical Relativistic Particle
\jour J. Appl. Prob. \vol 4\pages 389--396\yr1967\endref
\ref\key 22 \by Goldstein, H.\book Classical Mechanics
\publ Addison-Wesley Inc. \yr 1980\endref
\ref\key 23\by Jammer, M., \book The Philosophy of Quantum Mechanics,
John Wiley and Sons \publaddr N.Y. \yr1974\endref
\ref\key 24 \by J\'anossy, L.\jour Foundations of Physics \vol 6\pages 341\yr 1976\endref
\ref\key 25 \by Kershaw, D.\jour Phys. Rev. \vol 136\pages 1850\yr 1964\endref
\ref\key 26\by Madelung, E.\jour Zeits. f\"ur. Phys\vol 40\pages 332\yr 1926\endref
\ref\key 27\by Milne-Thomson, L. M.\book Theoretical Hydrodynamics
\publ Macmillan and Co Ltd\yr 1954\endref
\ref\key 28\by Nelson, E.\jour Phys. Rev. \vol 150\pages 1079\yr 1966\endref
\ref\key 29\by Peres, A.\jour Am. J. Phys.\vol 54\issue 8, August, \yr 1986\endref
\ref\key 30\by Takabayasi, T. \jour Progress of Theor. Phys.\vol 9\pages 187\yr 1953\endref
\ref\key 31 \by Thirring, W.E.\book Principles of Quantum Electrodynamics
\publ Academic Press Inc. \yr 1958\endref
\ref\key 32\by Toll, J.S.\jour Phys.Rev. \vol104\pages 1760\yr1956\endref
\ref\key 33\by Gilson, J.G.\jour Speculations in Science and Technology
\vol 27\issue 3 \pages 201-204\yr1994\endref
\ref\key 34 \by Cohen, E. R. and Taylor, B. N.\paper The Fundamental Physical Constants
\jour Physics Today (August)\pages BG9-15\yr 1993\endref


\endRefs
\enddocument